\documentclass[floatfix,showpacs,superscriptaddress,nofootinbib,twocolumn,showkeys]{revtex4-2}

\usepackage[T1]{fontenc}
\usepackage{graphicx}
\usepackage{xcolor}
\usepackage{amsmath}
\usepackage{hyperref}

\usepackage{tikz}
\usetikzlibrary{positioning}

\hypersetup{
    colorlinks=true,
    linkcolor=blue,
    filecolor=magenta,      
    urlcolor=blue,
    citecolor=blue
}

\usepackage{amssymb}
\usepackage{multirow,array}
\usepackage{lipsum}
\usepackage{enumitem}

\newcommand{\be}{\begin{equation}}
\newcommand{\ee}{\end{equation}}
\newcommand{\bs}{\begin{subequations}}
\newcommand{\es}{\end{subequations}}
\newcommand{\beal}{\begin{align}}

\newcommand{\trento}{\texttt{T$_\mathtt{R}$ENTo}}

\newcommand{\music}{\texttt{MUSIC}}

\newcommand{\sqrts}{\sqrt{s_\textrm{NN}}}
\newcommand{\Tsw}{T_{\text{sw}}}

\newcommand{\vecr}{\boldsymbol r}
\newcommand{\vecq}{\boldsymbol q}
\newcommand{\vecp}{\boldsymbol p}
\newcommand{\vecxi}{\boldsymbol \xi}

\begin{document}
\title{Role of bulk viscosity in deuteron production in ultrarelativistic nuclear collisions}
\date{\today}

\author{D.~Everett}
\affiliation{Department of Physics, The Ohio State University, Columbus OH 43210.}

\author{D.~Oliinychenko}
\affiliation{Institute for Nuclear Theory, Department of Physics, University of Washington, Seattle, WA 98195.}
\affiliation{Nuclear Science Division, Lawrence Berkeley National Laboratory, Berkeley CA 94270.}

\author{M.~Luzum}
\affiliation{Instituto  de  F\`{i}sica,  Universidade  de  S\~{a}o  Paulo,  C.P.  66318,  05315-970  S\~{a}o  Paulo,  SP,  Brazil. }

\author{J.-F. Paquet}
\affiliation{Department of Physics, Duke University, Durham NC 27708.}

\author{G.~Vujanovic}
\affiliation{Department of Physics and Astronomy, Wayne State University, Detroit MI 48201.}

\author{S.~A.~Bass}
\affiliation{Department of Physics, Duke University, Durham NC 27708.}

\author{L.~Du}
\affiliation{Department of Physics, McGill University, Montr\'{e}al QC H3A\,2T8, Canada.}

\author{C.~Gale}
\affiliation{Department of Physics, McGill University, Montr\'{e}al QC H3A\,2T8, Canada.}

\author{M.~Heffernan}
\affiliation{Department of Physics, McGill University, Montr\'{e}al QC H3A\,2T8, Canada.}

\author{U.~Heinz}
\affiliation{Department of Physics, The Ohio State University, Columbus OH 43210.}

\author{L.~Kasper}
\affiliation{Department of Physics and Astronomy, Vanderbilt University, Nashville TN 37235.}

\author{W.~Ke}
\affiliation{Los Alamos National Laboratory, Theoretical Division, Los Alamos, NM 87545.}
\affiliation{Department of Physics, University of California, Berkeley CA 94270.}
\affiliation{Nuclear Science Division, Lawrence Berkeley National Laboratory, Berkeley CA 94270.}

\author{D.~Liyanage}
\affiliation{Department of Physics, The Ohio State University, Columbus OH 43210.}

\author{A.~Majumder}
\affiliation{Department of Physics and Astronomy, Wayne State University, Detroit MI 48201.}

\author{A.~Mankolli}
\affiliation{Department of Physics and Astronomy, Vanderbilt University, Nashville TN 37235.}

\author{C.~Shen}
\affiliation{Department of Physics and Astronomy, Wayne State University, Detroit MI 48201.}
\affiliation{RIKEN BNL Research Center, Brookhaven National Laboratory, Upton NY 11973.}

\author{D.~Soeder}
\affiliation{Department of Physics, Duke University, Durham NC 27708.}

\author{J.~Velkovska}
\affiliation{Department of Physics and Astronomy, Vanderbilt University, Nashville TN 37235.}


\author{A.~Angerami}
\affiliation{Lawrence Livermore National Laboratory, Livermore CA 94550.}

\author{R.~Arora}
\affiliation{Research Computing Group, University Technology Solutions, The University of Texas at San Antonio, San Antonio TX 78249.}

\author{S.~Cao}
\affiliation{Institute of Frontier and Interdisciplinary Science, Shandong University, Qingdao, Shandong 266237, China}

\author{Y.~Chen}
\affiliation{Laboratory for Nuclear Science, Massachusetts Institute of Technology, Cambridge MA 02139.}
\affiliation{Department of Physics, Massachusetts Institute of Technology, Cambridge MA 02139.}



\author{T.~Dai}
\affiliation{Department of Physics, Duke University, Durham NC 27708.}

\author{R.~Ehlers}
\affiliation{Department of Physics and Astronomy, University of Tennessee, Knoxville TN 37996.}
\affiliation{Physics Division, Oak Ridge National Laboratory, Oak Ridge TN 37830.}
\affiliation{Department of Physics, University of California, Berkeley CA 94270.}
\affiliation{Nuclear Science Division, Lawrence Berkeley National Laboratory, Berkeley CA 94270.}

\author{H.~Elfner}
\affiliation{GSI Helmholtzzentrum f\"{u}r Schwerionenforschung, 64291 Darmstadt, Germany.}
\affiliation{Institute for Theoretical Physics, Goethe University, 60438 Frankfurt am Main, Germany.}
\affiliation{Frankfurt Institute for Advanced Studies, 60438 Frankfurt am Main, Germany.}

\author{W.~Fan}
\affiliation{Department of Physics, Duke University, Durham NC 27708.}

\author{R.~J.~Fries}
\affiliation{Cyclotron Institute, Texas A\&M University, College Station TX 77843.}
\affiliation{Department of Physics and Astronomy, Texas A\&M University, College Station TX 77843.}

\author{F.~Garza}
\affiliation{Cyclotron Institute, Texas A\&M University, College Station TX 77843.}
\affiliation{Department of Physics and Astronomy, Texas A\&M University, College Station TX 77843.}

\author{Y.~He}
\affiliation{Key Laboratory of Quark and Lepton Physics (MOE) and Institute of Particle Physics, Central China Normal University, Wuhan 430079, China.}
\affiliation{Guangdong Provincial Key Laboratory of Nuclear Science, Institute of Quantum Matter, South China Normal University, Guangzhou 510006, China}
\affiliation{Guangdong-Hong Kong Joint Laboratory of Quantum Matter, Southern Nuclear Science Computing Center, South China Normal University, Guangzhou 510006, China}

\author{B.~V.~Jacak}
\affiliation{Department of Physics, University of California, Berkeley CA 94270.}
\affiliation{Nuclear Science Division, Lawrence Berkeley National Laboratory, Berkeley CA 94270.}

\author{P.~M.~Jacobs}
\affiliation{Department of Physics, University of California, Berkeley CA 94270.}
\affiliation{Nuclear Science Division, Lawrence Berkeley National Laboratory, Berkeley CA 94270.}

\author{S.~Jeon}
\affiliation{Department of Physics, McGill University, Montr\'{e}al QC H3A\,2T8, Canada.}


\author{M.~Kelsey}
\affiliation{Department of Physics and Astronomy, Wayne State University, Detroit MI 48201.}


\author{M.~Kordell~II}
\affiliation{Cyclotron Institute, Texas A\&M University, College Station TX 77843.}
\affiliation{Department of Physics and Astronomy, Texas A\&M University, College Station TX 77843.}

\author{A.~Kumar}
\affiliation{Department of Physics, McGill University, Montr\'{e}al QC H3A\,2T8, Canada.}
\affiliation{Department of Physics and Astronomy, Wayne State University, Detroit MI 48201.}

\author{J.~Latessa}
\affiliation{Department of Computer Science, Wayne State University, Detroit MI 48202.}

\author{Y.-J.~Lee}
\affiliation{Laboratory for Nuclear Science, Massachusetts Institute of Technology, Cambridge MA 02139.}
\affiliation{Department of Physics, Massachusetts Institute of Technology, Cambridge MA 02139.}

\author{A.~Lopez}
\affiliation{Instituto  de  F\`{i}sica,  Universidade  de  S\~{a}o  Paulo,  C.P.  66318,  05315-970  S\~{a}o  Paulo,  SP,  Brazil. }


\author{S.~Mak}
\affiliation{Department of Statistical Science, Duke University, Durham NC 27708.}

\author{C.~Martin}
\affiliation{Department of Physics and Astronomy, University of Tennessee, Knoxville TN 37996.}

\author{H.~Mehryar}
\affiliation{Department of Computer Science, Wayne State University, Detroit MI 48202.}


\author{T.~Mengel}
\affiliation{Department of Physics and Astronomy, University of Tennessee, Knoxville TN 37996.}

\author{J.~Mulligan}
\affiliation{Department of Physics, University of California, Berkeley CA 94270.}
\affiliation{Nuclear Science Division, Lawrence Berkeley National Laboratory, Berkeley CA 94270.}

\author{C.~Nattrass}
\affiliation{Department of Physics and Astronomy, University of Tennessee, Knoxville TN 37996.}




\author{J.~H.~Putschke}
\affiliation{Department of Physics and Astronomy, Wayne State University, Detroit MI 48201.}

\author{G.~Roland}
\affiliation{Laboratory for Nuclear Science, Massachusetts Institute of Technology, Cambridge MA 02139.}
\affiliation{Department of Physics, Massachusetts Institute of Technology, Cambridge MA 02139.}

\author{B.~Schenke}
\affiliation{Physics Department, Brookhaven National Laboratory, Upton NY 11973.}

\author{L.~Schwiebert}
\affiliation{Department of Computer Science, Wayne State University, Detroit MI 48202.}

\author{A.~Silva}
\affiliation{Department of Physics and Astronomy, University of Tennessee, Knoxville TN 37996.}

\author{C.~Sirimanna}
\affiliation{Department of Physics and Astronomy, Wayne State University, Detroit MI 48201.}

\author{R.~A.~Soltz}
\affiliation{Department of Physics and Astronomy, Wayne State University, Detroit MI 48201.}
\affiliation{Lawrence Livermore National Laboratory, Livermore CA 94550.}

\author{J.~Staudenmaier}
\affiliation{GSI Helmholtzzentrum f\"{u}r Schwerionenforschung, 64291 Darmstadt, Germany.}

\author{M.~Strickland}
\affiliation{Department of Physics, Kent State University, Kent, OH 44242.}

\author{Y.~Tachibana}
\affiliation{Akita International University, Yuwa, Akita-city 010-1292, Japan.}

\author{X.-N.~Wang}
\affiliation{Department of Physics, University of California, Berkeley CA 94270.}
\affiliation{Nuclear Science Division, Lawrence Berkeley National Laboratory, Berkeley CA 94270.}

\author{R.~L.~Wolpert}
\affiliation{Department of Statistical Science, Duke University, Durham NC 27708.}



\collaboration{The JETSCAPE Collaboration}

\begin{abstract}
    We use a Bayesian-calibrated multistage viscous hydrodynamic  model to explore deuteron yield, mean transverse momentum and flow observables in LHC Pb-Pb collisions. We explore theoretical uncertainty in the production of deuterons, including (i) the contribution of thermal deuterons, (ii)  models for the subsequent formation of deuterons (hadronic transport vs coalescence) and (iii) the overall sensitivity of the results to the hydrodynamic model --- in particular to bulk viscosity, which is often neglected in studies of deuteron production. Using physical parameters set by a comparison to only light hadron observables, we find good agreement with measurements of the mean transverse momentum $\langle p_T \rangle$ and  elliptic flow $v_2$ of deuterons; however, tension is observed with experimental data for the deuteron multiplicity in central collisions. The results are found to be sensitive to each of the mentioned theoretical uncertainties, with a particular sensitivity to bulk viscosity, indicating that the latter is an important ingredient for an accurate treatment of deuteron production.
\end{abstract}

\maketitle

\section{Introduction}
\label{sec:intro}

Ultra-relativistic ion collisions produce a hot strongly-coupled plasma of quarks and gluons which expands, cools down and recombines into pions, nucleons, other hadrons, and light nuclei. The recent measurements of deuterons, $^3\mathrm{He}$, $^4\mathrm{He}$, and $^3_{\Lambda}\mathrm{H}$ at the Large Hadron Collider (LHC) in Pb-Pb collisions at $\sqrts = $ 2.76 TeV and 5.02 TeV \cite{Adam:2015vda,Adam:2015yta,Acharya:2017bso,ALICE:2020chv} have renewed interest in the mechanisms of light nucleus production. The improved precision of measurements provides an opportunity to revisit and test current models of deuteron production.

Three broad types of models are typically used to calculate the production of deuterons in such collisions. The thermal model assumes that light nuclei reach equilibrium with hadrons, until a point of chemical freeze-out, where hadron and nuclear abundances are frozen, which happens at approximately the same temperature across the entire system~\cite{Andronic:2017pug}. To obtain a momentum distribution for the light nuclei, one can combine the thermal model with a blast wave model at a later \emph{kinetic} freeze-out \cite{Bellini:2018epz}. Another approach is the coalescence model, which predicts that the number of produced light nuclei is a convolution of (i) a source function characterizing the nucleons' distribution in phase space, and (ii) the light nucleus' Wigner function.\footnote{%
    Although the quantum mechanical foundations of the coalescence model have been studied for many years \cite{Butler:1961pr, Sato:1981ez, Scheibl:1998tk, Blum:2019suo}, modern implementations can still vary substantially both in methods and results, especially for nuclei heavier than the deuteron. Different types of coalescence models are listed, for example, in Ref.~\cite{Oliinychenko:2020ply}. We will use a model where nucleons from transport, free-streamed to the larger of their last interaction times, are convoluted with a Gaussian deuteron Wigner function in their center of mass frame; see Section~\ref{sec:theory_deuteron_coalescence}.}
Realistic source functions (space-time-momentum distributions) for the final hadrons can be obtained from  hydrodynamics and transport models of heavy-ion collisions. The main assumption of the coalescence model is that light nuclei are formed 
when hadronic interactions cease. Finally, one can model the production of light nuclei with a transport approach which implements specific production and dissociation reactions --- for example $N p n \leftrightarrow N d$, $\pi p n \leftrightarrow \pi d$, or $\pi d \leftrightarrow p n$ \cite{Danielewicz:1991dh, Oh:2009gx, Oliinychenko:2018ugs, Oliinychenko:2018odl, Oliinychenko:2020znl} ($N$ denotes either $p$ or $n$) --- or binding of nucleons in transport by potentials \cite{Kireyeu:2021igi, glassel2021cluster}. 

Ultra-relativistic heavy-ion collisions are frequently simulated by multistage approaches, where the evolution of the plasma is modeled by relativistic viscous hydrodynamics, and the subsequent hadronic rescattering at a more dilute stage is described by kinetic theory. Conveniently, in a multistage simulation, one can test all three types of light nucleus production models. One can sample light nuclei from a near-equilibrium distribution at the transition from hydrodynamics to transport, similar to a thermal model at chemical freeze out; one can include the reactions involving deuterons into the transport phase; or one can use the final state nucleons as a source function for coalescence. 

In all of these models of light nucleus production, there is a close connection between the distribution of nucleons and the light nucleus bound states that they form.  Therefore, it is evident that a successful description of light nuclei relies on a good description of the underlying system evolution.

To test the above light nucleus production models, we employ a hybrid hydrodynamic and hadronic transport model that was calibrated to a wide set of hadron observables~\cite{JETSCAPE:2020shq, JETSCAPE:2020mzn}. Importantly, this multistage model includes bulk viscosity, unlike most previous studies of light nuclei.
While uncertainties remain in the magnitude of bulk viscosity of QCD, and in particular the modeling of its effect on particlization, it is generically expected to have a larger effect on heavier particles, and therefore should be important for the production of light nuclei. We show in this work that bulk viscosity indeed has a large effect on deuteron production, regardless of the underlying deuteron production model.

This work is organized as follows. In Section~\ref{sec:model_overview}, we first briefly describe our multistage hydrodynamic model and deuteron production models used in this work. 
We then compare the multistage model with measured deuteron observables --- yield $dN/dy$, mean transverse momenta $\langle p_T \rangle$, and azimuthal angular anisotropy $v_2$ --- in Pb-Pb at the LHC in Sections~\ref{sec:model_comparisons}-\ref{sec:spectra_flow}.
We further quantify via Bayesian inference the additional constraints provided by deuteron observables on the initial conditions and bulk viscosity of the plasma in Section~\ref{sec:sensitivity_bulk}, and summarize our results in Section~\ref{sec:conclusions}.

\section{Model Overview}
\label{sec:model_overview}
\subsection{Multistage model of heavy ion collisions}
\label{sec2a}

A detailed description of the hybrid hydrodynamic-transport model used throughout this work can be found in Ref.~\cite{JETSCAPE:2020mzn}. Briefly, the \trento{} model~\cite{Moreland:2014oya} is used as a parametric initialization for the energy density shortly after the impact of the nuclei. This profile is then free-streamed for a short proper time, and is then used as initialization for second-order relativistic viscous hydrodynamics (\music{}~\cite{Schenke:2010nt, Schenke:2010rr, Paquet:2015lta}). We use the shear and bulk viscosity parametrizations described in Refs.~\cite{JETSCAPE:2020mzn, JETSCAPE:2020shq} and an equation of state which matches the trace anomaly of lattice calculations~\cite{Bazavov:2014pvz} and the hadron resonance gas. On a surface of constant temperature $\Tsw{}$, the Cooper-Frye prescription \cite{Cooper:1974qi} is employed to switch description (``particlization'') from fluid to a kinetic theory of hadrons, which then scatter, decay and form resonances in the SMASH hadronic afterburner \cite{Weil:2016zrk, smash_code, dmytro_oliinychenko_2019_3485108}. 

In the ideal hydrodynamic limit, as in the case of thermal models, the hadrons in the fluid are in local equilibrium. In kinetic theory, it implies that their momentum distribution is given by the Bose-Einstein or Fermi-Dirac distribution. In the more general case where there is dissipation in hydrodynamics, there is an off-equilibrium correction to the hadron distribution, and this viscous correction is model dependent \cite{Damodaran:2020qxx, McNelis:2021acu, Molnar:2014fva}. In this work, we use the ``Grad'' model~\cite{CPA:CPA3160020403} for viscous corrections to the equilibrium distribution function, the one out of several studied in Refs.~\cite{JETSCAPE:2020mzn,JETSCAPE:2020shq} that gave the best agreement with light hadron measurements (listed below).

The model parameters were calibrated by Bayesian parameter estimation against observables for Pb-Pb collisions at $\sqrts{}=2.76$ TeV as well as Au-Au collisions at $\sqrts{}=0.2$ TeV. The Pb-Pb observables include the yields and mean transverse momenta of pions, kaons and protons~\cite{ALICE:2013mez}, the charged particle multiplicity and transverse energy~\cite{Aamodt:2010cz, Adam:2016thv}, the charged particle elliptic, triangular and quadrangular flow~\cite{ALICE:2011ab}, and the charged particle mean transverse momentum fluctuations \cite{Abelev:2014ckr}.  The calibration observables for Au-Au collisions include the yields and mean transverse momenta of pions and kaons~\cite{Abelev:2008ab}, and the charged particle elliptic and triangular flow~\cite{Adams:2004bi, Adamczyk:2013waa}. In particular, no deuteron or light nuclei measurements were used in the calibration of these parameters. The parameters used to generate the predictions in the next section are listed in Table~\ref{tab:table_params}.\footnote{%
    The parameters in Table~\ref{tab:table_params} are slightly different from the Maximum A Posteriori parameters reported in Refs.~\cite{JETSCAPE:2020mzn, JETSCAPE:2020shq}, but remain parameters of high posterior probability density. The multidimensional posterior does not have a very sharply defined global maximum, and the Maximum A Posteriori parameters can differ slightly depending on the details of the Markov Chain Monte Carlo optimization.}

%
%

\begin{table}[b]
\centering
\begin{tabular}{l|r||l|r||l|r}
\hline
parameter & value & parameter & value & parameter & value\\ \hline\hline
$N$(2.76 TeV)[GeV] & 14.2  &$\tau_R$[fm/c] &  1.48 & $(\zeta/s)_{\text{max}}$ & 0.13\\ \hline
$N$(5.02 TeV)[GeV] & 18.8  &$\alpha$ &  0.047 & $T_{\zeta, c}$[GeV] &  0.12 \\ \hline
$T_{\eta, \text{kink}}$[GeV] & 0.22 &$p$ &  0.06 & $w_{\zeta}$[GeV]  &  0.089\\ \hline
 $a_{\eta, \text{low}}$[GeV$^{-1}$]& $-0.76$   &$\sigma_k$ & 0.98 & $\lambda_{\zeta}$ & $-0.19$\\ \hline
 $a_{\eta, \text{high}}$[GeV$^{-1}$]&  0.22  &$w$[fm] &  1.12 & $b_{\pi}$ &  4.5\\ \hline
 $(\eta/s)_{\text{kink}}$ & 0.096  &$d_{\text{min}}^3$[fm$^3$] &  2.97  & $\Tsw{}$[GeV] & 0.136\\ \hline
\end{tabular}
\caption{Model parameters used to produce the predictions in this work. See Refs.~\cite{JETSCAPE:2020mzn,JETSCAPE:2020shq} for details.}
\label{tab:table_params}
\end{table}

We explore several different models of deuteron production, 
including sampling on the switching hypersurface, 
production via three-body scattering in the hadronic afterburner, 
and coalescence on the kinetic freezeout surface. 

\subsection{Deuteron production with transport}
\label{sec:theory_deuteron_transport}

We investigate two models of deuteron production with transport, 
which are (i): ``transport only'', which assumes that no deuterons are present at the transition from hydrodynamics to hadronic transport (i.e., at ``particlization''); rather, all deuterons are created during the hadronic rescattering phase by reactions --- $\pi d \leftrightarrow \pi np$, $N d \leftrightarrow N np$, $\bar{N} d \leftrightarrow \bar{N} np$, $\pi d \leftrightarrow NN$ and all of their CPT-conjugates, with elastic $\pi d$, $N d$ and $\bar{N} d$ also changing the deuterons' momenta; (ii): ``Cooper-Frye + transport'' which assumes that deuterons are nearly equilibrated with the hadron resonance gas at the transition from hydrodynamics, so that they are sampled according to near-equilibrium distribution functions and subsequently allowed to be formed and/or destroyed in the hadronic rescattering phase.

Notice that the yield of deuterons at the Cooper-Frye sampling in the model (ii) is closely related to the results of a thermal model (discussed in the introduction) with freeze-out temperature $\Tsw{}$ and volume $V = \int d\sigma_{\mu}u^{\mu}$,
where $d\sigma_{\mu}$ are the normal 4-vectors to the hypersurface, and $u^{\mu}$ are the collective velocities of the hypersurface elements.  These quantities enter the Cooper-Frye formula~\cite{Cooper:1974qi}
\begin{equation}
P^0 \frac{d N}{d^3p}=\frac{g}{(2\pi\hbar)^3} \int d\sigma_{\mu} P^{\mu} \left( f_{eq}(P_{\nu} u^{\nu}/T) + \delta f \right) \,,
\end{equation}
with $f(P_{\nu} u^{\nu}/T)$ for deuterons being the Bose-Einstein distribution, $\delta f$ being a correction due to viscosity, and $g$ their degeneracy factor. Typically, thermal models assume local equilibrium, $\delta f = 0$.

In this work we use the implementation of these $3 \to 2$ reactions via an intermediate fictitious $d'$ resonance~\cite{Oliinychenko:2018ugs}: $pn \leftrightarrow d'$, $\pi d' \leftrightarrow \pi d$, $N d' \leftrightarrow N d$. Recently a possibility to simulate $3 \to 2$ reactions directly, without $d'$, via stochastic rates was implemented in \texttt{SMASH} ~\cite{Staudenmaier:2021lrg}, but it is not employed in this work. In Ref.~\cite{Staudenmaier:2021lrg} it was shown that the main difference between  direct $3\to 2$ reactions and reactions with the intermediate $d'$ resonance is that $3\to 2$ reactions bring the deuteron yield to equilibrium faster.

\subsection{Deuteron production with coalescence}
\label{sec:theory_deuteron_coalescence}

Here we employ a Wigner-function coalescence model~\cite{Scheibl:1998tk} that has no free parameters. The implementation of coalescence is based on the equation
\begin{multline} \label{eq:coalescence1}
    \frac{d^3N_d}{d^3\vecp_d} = \frac{3}{8} \int \frac{d^3\vecr_d d^3\vecr  d^3\vecq}{(2\pi\hbar)^6} \mathcal{D}(\vecr,\vecq) \times  \\ W\left(\frac{\vecp_d}{2} + \vecq, \frac{\vecp_d}{2} - \vecq, \vecr_d + \frac{\vecr}{2}, \vecr_d - \frac{\vecr}{2} \right) \,,
\end{multline}
where the factor $3/8$ originates from spin and isospin averaging. The deuteron Wigner function is $\mathcal{D}(\vecr,\vecq) = 8 \exp(-|\vecr|^2/d^2 - |\vecq|^2 d^2 / \hbar^2) $, which originates from
\begin{align}
    \mathcal{D}(\vecr,\vecq) = \int d^3\vecxi e^{i \vecq \cdot \vecxi/\hbar} \varphi_d(\vecr + \vecxi/2)\varphi_d^*(\vecr - \vecxi/2)\,,
\end{align}
where the deuteron wavefunction $\varphi_d$ is assumed to be a Gaussian with $d = 3.2$ fm~\cite{Kachelriess:2020amp}. The function $W$ is the probability to find a pair of nucleons at positions $\vecr_d \pm \vecr/2$ with momenta $\vecp_d/2 \pm \vecq$. In our model the distribution $W$ is represented by the nucleon pairs in the hadronic afterburner. The integrals in Eq.~(\ref{eq:coalescence1}) are computed as a sum over all nucleon pairs in the simulation. Every pair is transported in time to the latest of their last collision times. At this moment $\vecr_1$, $\vecr_2$ are coordinates and $\vecp_1$ and $\vecp_2$ are momenta of the nucleons. Then
\begin{align}
    \vecr_d &= \frac{1}{2} (\vecr_1 + \vecr_2)\,, \quad
    \vecr = \vecr_1 - \vecr_2\,, \\
    \vecp_d &= \vecp_1 + \vecp_2\,,\quad
    \vecq = \frac{1}{2}(\vecp_1 - \vecp_2)\,,
\end{align}
and the pair contributes to deuteron spectra with the weight $\frac{3}{8}\mathcal{D}(\vecr,\vecq)$.

\section{Results}
\label{sec3}
\subsection{Comparing transport, coalescence, and thermal-like deuteron production}
\label{sec:model_comparisons}

To investigate the different mechanisms of deuteron production, we compare the three specific scenarios presented in Sections~\ref{sec:theory_deuteron_transport} and \ref{sec:theory_deuteron_coalescence} within our multistage model of heavy ion collisions:
\begin{enumerate}
\item[(i)] Transport only: deuterons not present at particlization, followed by hadronic transport with deuteron reactions;
\item[(ii)] Cooper-Frye + Transport: deuterons present at particlization and allowed to react in the hadronic transport;
\item[(iii)] \label{test}
 Coalescence only: deuterons not present at particlization, hadronic transport without deuteron reactions, followed by coalescence at kinetic freeze-out.
\end{enumerate}

While these do not represent all possible variations, they provide sufficient information for understanding the effects of each mechanism.

\begin{figure}[tb]
  \begin{tikzpicture}
    \node (img1) {\includegraphics[width=0.9\linewidth]{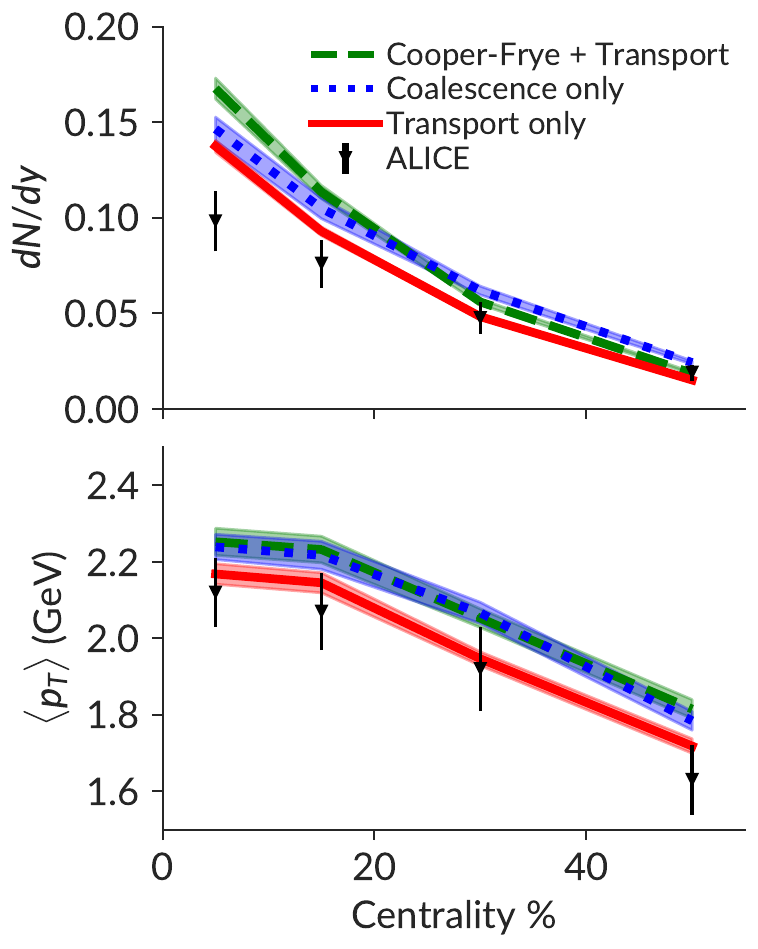}};
    \node[above right = -6cm and -2.5cm of img1] (img2) {\includegraphics[width=0.2\linewidth]{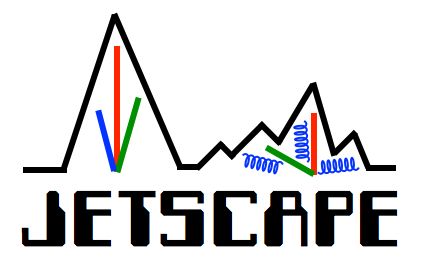}};
    \end{tikzpicture}
  \caption{Comparison of the deuteron midrapidity yield (top) and mean transverse momentum (bottom) in $\sqrts{}=2.76$~TeV Pb-Pb collisions for three scenarios of deuteron production. Experimental data from Ref.~\cite{ALICE:2015wav}. See text for more details.}
  \label{fig:3-way}
\end{figure}

In Fig.~\ref{fig:3-way} one can see that these three scenarios result in rather similar yield and mean transverse momentum of deuterons. Let us concentrate on the comparison of the transport scenarios first. As in previous studies \cite{Oliinychenko:2018odl,Oliinychenko:2018ugs}, deuteron yields and average transverse momenta are similar whether or not deuterons are sampled at particlization. To understand this effect, note that (i) the reaction $\pi d \rightarrow \pi p n$ has a large cross section, (ii) the implementation of the reverse reaction obeys the detailed balance principle, and (iii) a large number of pions is produced in heavy ion collisions. Given these conditions, as long as the $\pi d \leftrightarrow \pi p n$ reaction rate is larger than the expansion rate of the plasma, deuterons rapidly approach relative equilibrium with protons. When the deuteron is not sampled at particlization,  the deuteron yield approaches but does not completely reach equilibrium \cite{Oliinychenko:2018odl, Oliinychenko:2018ugs}, because the fireball's expansion freezes out the $\pi d \leftrightarrow \pi p n$ reactions; this is the reason for the slightly smaller number of deuterons found in the ``Transport only'' scenario. Note that, to avoid confusion, one should think about deuterons being in equilibrium in a statistical sense, averaging over a large number of heavy-ion collisions, and not in a single event --- notice that on average only around 0.1 deuterons are produced per event per unit of rapidity, as shown in Fig.~\ref{fig:3-way}.

In Refs.~\cite{Oliinychenko:2018odl,Oliinychenko:2018ugs} both scenarios (i) and (ii) were found to be compatible with the experimental data, although sampling deuterons at particlization was slightly preferred. In this work, the situation is different: the experimentally measured deuteron yield and transverse momentum are better described by omitting the deuteron from particlization (see Fig. \ref{fig:3-way}). While there are multiple differences between the multistage models used in the present work and Refs.~\cite{Oliinychenko:2018odl, Oliinychenko:2018ugs}, the likely reason for this difference is bulk viscosity, which was not included in previous works. We discuss this effect in detail in Section~\ref{sec:sensitivity_bulk}. We note that the (mostly systematic) uncertainties are still significant on the deuteron measurements~\cite{ALICE:2015wav} shown in Fig.~\ref{fig:3-way}. Reduction of these systematic uncertainties would increase the discrimination power of deuterons even further.

The creation of deuterons by reactions has a certain similarity with coalescence. With reactions, most of final-state deuterons are produced rather late; earlier produced deuterons tend to get destroyed by subsequent collisions. Moreover, by the kinematics of the $\pi p n \to \pi d$ reaction, the incoming proton and neutron have to be close in phase space, as coalescence assumes. In Fig.~\ref{fig:3-way} one can indeed see that the coalescence model provides a very similar deuteron yield and transverse momentum as the models with deuteron-producing reactions. Therefore, it is the underlying proton phase space distribution (which is the same in all three models) that influences the deuteron observables here, while the deuteron production mechanism is less important.\footnote{%
    This may be different in small systems where the span of the deuteron wave function is comparable to the system size.} 
As a consequence, the model parameters that influence proton production will also influence deuteron production. This means that by combining proton and deuteron observables one could potentially constrain these parameters better than by using only proton observables.

Based on the agreement of model predictions with measured proton data, one might expect a better agreement for deuteron yields. There are several reasons why the deuteron yield is not as well described as expected. First, although the model is tuned to describe integrated proton yield and $\langle p_T \rangle$ precisely, the proton $p_T$-spectrum in fact exhibits deviations from experiment \cite[Fig.~17]{JETSCAPE:2020mzn}. Second, previous studies suggesting that a good agreement of proton data implies good agreement with deuteron data \cite{Oliinychenko:2018odl, Oliinychenko:2018ugs} did not take into account bulk viscosity, unlike in the present work. Importantly, the bulk viscous corrections to the proton $p_T$-spectrum are substantial (see Fig.~\ref{fig:MAP} below).

\subsection{Yield, mean $p_T$ and flow of deuterons at $\sqrts{}=$5.02 TeV}
\label{sec:spectra_flow}

In this section we provide a prediction for the multiplicity and mean $p_T$ of deuterons for Pb-Pb $\sqrts{}=$ 5.02 TeV collisions using the ``Transport only'' approach. We further compare our calculation for $p_T$-differential $v_2$ with ALICE measurements for deuterons in both Pb-Pb $\sqrts{}=$ 2.76 TeV and $\sqrts{}=$ 5.02 TeV. 

The initial conditions, transport coefficients and other parameters of the multistage model given in Table~\ref{tab:table_params} were not calibrated using Pb-Pb $\sqrts{} = 5.02$ TeV observables, only to Pb-Pb $\sqrts{} = 2.76$ TeV and Au-Au $\sqrts{} = 0.2$ TeV observables~\cite{JETSCAPE:2020mzn,JETSCAPE:2020shq}. We assume that all model parameters remain the same except for two \trento{} initial condition parameters that are expected to be center-of-mass energy dependent: (i) the nucleon-nucleon inelastic cross section and (ii)  the overall normalization of the initial energy density. For the inelastic nucleon-nucleon cross section at $\sqrts{} = 5.02$ TeV, we used 70~mb. The normalization of the initial energy density is typically a parameter tuned to heavy ion measurements, mostly the hadronic multiplicities. In our approach, instead of re-tuning it to measurements, we simply estimated it from a previous Bayesian inference that also used \trento{} initial condition~\cite{Bernhard:2019bmu}. In that work, they found the ratio of the normalization\footnote{%
    More specifically, these are the normalizations for the Maximum A Posteriori parameters of each system.} 
at  $\sqrts=$ 5.02 TeV and  $\sqrts=$ 2.76 TeV to be $1.32$, yielding the normalization value at  $\sqrts{} = 5.02$ TeV quoted in Table~\ref{tab:table_params}.

\begin{figure}[tb] 
\centering
  \begin{tikzpicture}
    \node (img1) {\includegraphics{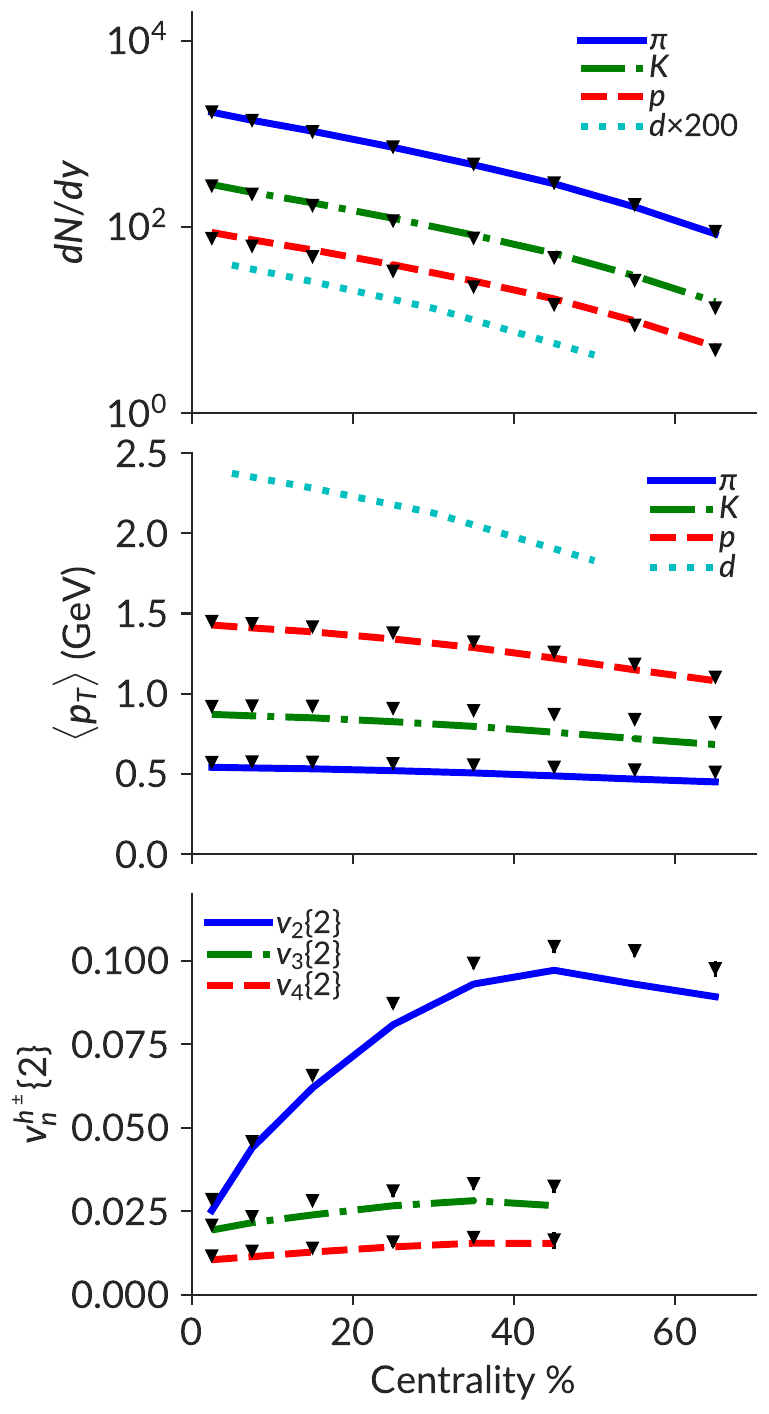}};
    \node[above right = -1.2cm and -5cm of img1] (img2) {\includegraphics[width=0.2\linewidth]{plots/jetscape_logo.jpg}};
    \end{tikzpicture}
\caption{Model predictions for Pb-Pb $\sqrts{} = 5.02$ TeV collisions with parameters given in Table \ref{tab:table_params}. The model was calibrated only to Pb-Pb $\sqrts{} = 2.76$ TeV and Au-Au $\sqrts{} = 0.2$~TeV observables. The deuteron was not sampled on the switching surface, but only allowed to form during the SMASH hadronic cascade (the ``Transport only'' scenario). ALICE measurements \cite{ALICE:2019hno, ALICE:2015juo, ALICE:2016ccg} are plotted as black triangles.
}
\label{fig:not_sample_d_5.02}
\end{figure}

\begin{figure}[tb]
  \centering
    \begin{tikzpicture}
    \node (img1) {    \includegraphics[width=8.3cm]{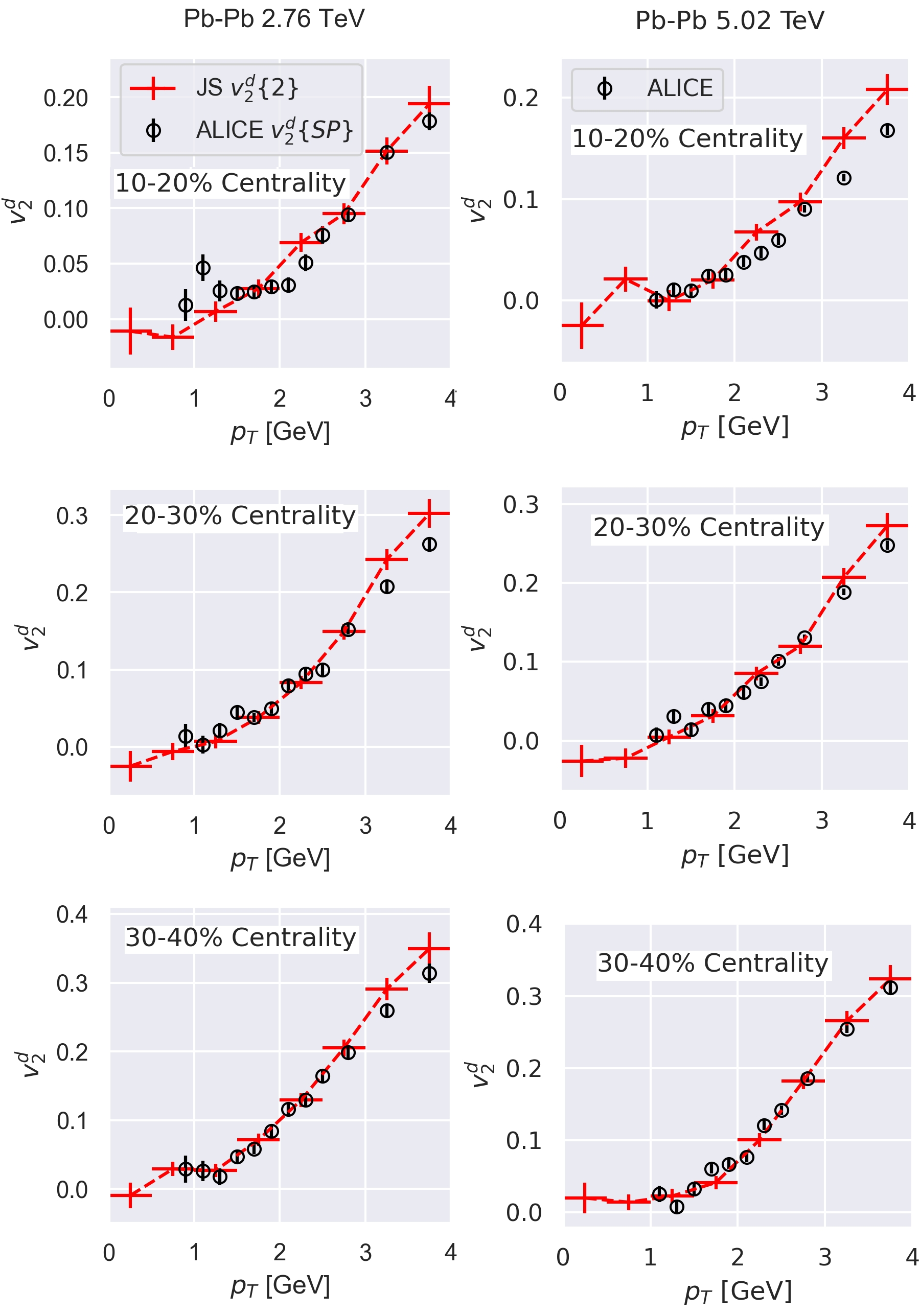}};
    \node[above right = -6.1cm and -3.2cm of img1] (img2) {\includegraphics[width=0.15\linewidth]{plots/jetscape_logo.jpg}};
    \end{tikzpicture}
  \caption{The differential $v_2$ for deuterons for three centrality bins for Pb-Pb collisions at 2.76 TeV (left) and 5.02 TeV (right). Our calculations for the ``transport only'' approach are red crosses, and ALICE measurements \cite{ALICE:2020chv} are black circles. The model observables are averaged over five thousand fluctuating initial conditions. 
  }
  \label{fig:diff_v2_d}
\end{figure}

The comparison with measurements from ALICE \cite{ALICE:2019hno, ALICE:2015juo, ALICE:2016ccg} is shown in Fig.~\ref{fig:not_sample_d_5.02}. The agreement between the calculations and hadron measurements is very similar to that found at $\sqrts{}=2.76$~TeV, see Ref.~\cite[Fig.~8]{JETSCAPE:2020mzn}. Our prediction for the deuteron multiplicity and mean $p_T$ at $\sqrts{}=5.02$~TeV is shown on the same figure. As is the case for light hadrons, it is natural to expect our predictions for deuterons at 5.02 TeV to have very similar agreement as for the 2.76 GeV results (see the ``Transport only'' curve in Fig.~\ref{fig:3-way}) --- that is, generally good agreement except for an overestimated yield in central collisions.

The $p_T$-differential $v_2$ of deuterons in Pb-Pb collisions at $\sqrts{} = $ 2.76 TeV and  5.02 TeV is described very well for different collision centralities, as shown in Fig.~\ref{fig:diff_v2_d}. The $\sqrts{}=$ 5.02 TeV $v_2$ result was shown as a prediction in the ALICE publication~\cite{ALICE:2020chv}. We evaluate the $p_T$-differential deuteron momentum anisotropy $v_2\{2\}$ using the Q-cumulant method~\cite{Bilandzic:2010jr}.

\subsection{Sensitivity to medium properties}
\label{sec:sensitivity_bulk}

\begin{figure}[tb]
  \centering

    \begin{tikzpicture}
    \node (img1) {\includegraphics[width=0.45\textwidth]{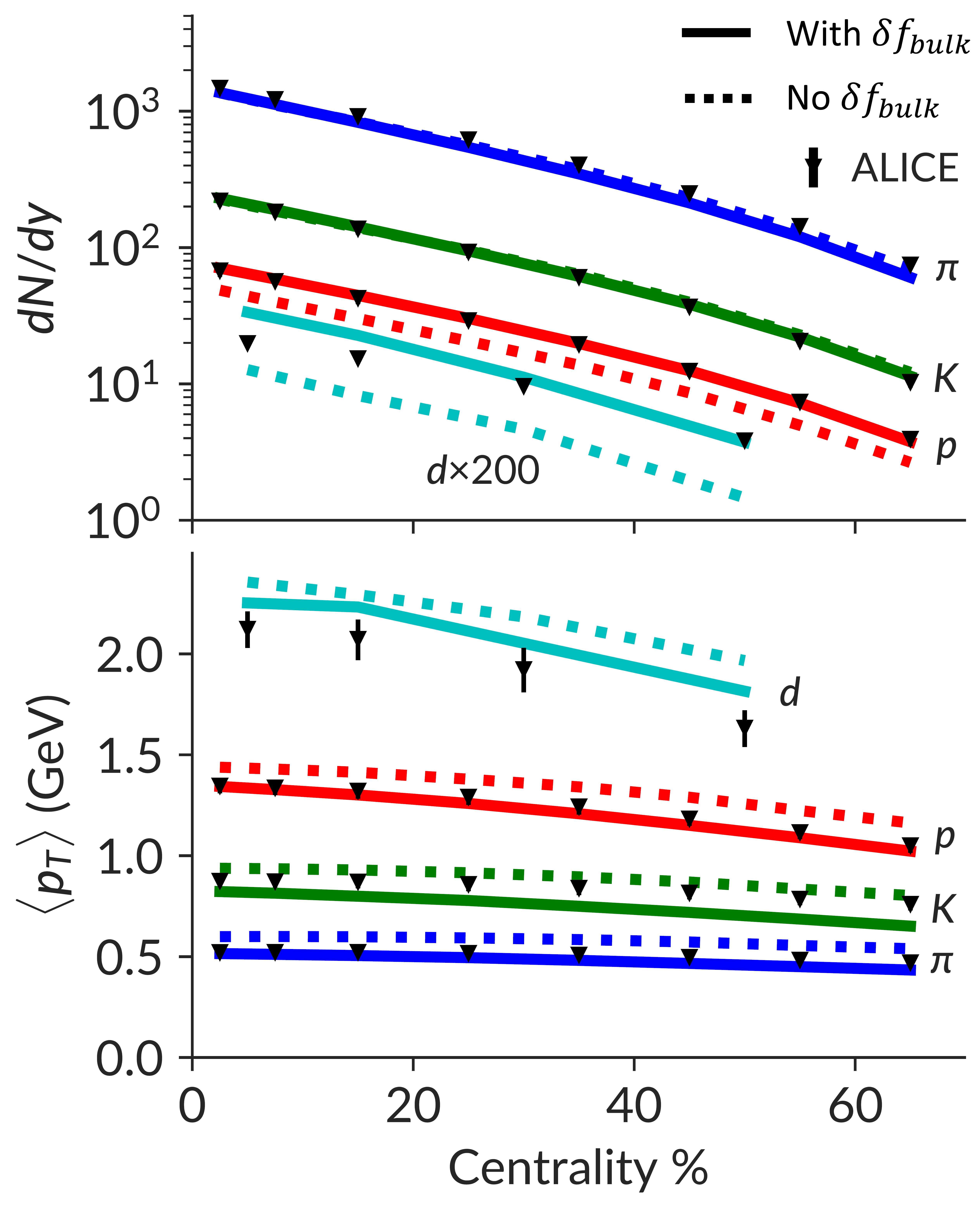}};
    \node[above right = -1.0cm and -5.4cm of img1] (img2) {\includegraphics[width=0.17\linewidth]{plots/jetscape_logo.jpg}};
    \end{tikzpicture}
    
  \caption{Identified hadron multiplicity (top) and mean $p_T$ (bottom) for Pb-Pb $\sqrts{}=2.76$~TeV as a function of centrality, with (solid line) or without (dotted line) the bulk viscous correction $\delta f_\textrm{bulk}$ in Cooper-Frye for all particles. Deuterons are produced at particlization and allowed to dynamically form and be destroyed, corresponding to the ``Cooper-Frye+Transport'' scenario discussed in Section~\ref{sec:model_comparisons}. Note that the effect of the viscous correction on pions, protons and other hadrons propagate to deuterons through the transport phase.
  ALICE measurements \cite{ALICE:2013mez,ALICE:2015wav} are plotted as black triangles.}
  \label{fig:MAP}
\end{figure}

In this section, we explore the sensitivity of the deuteron yield and mean transverse momentum to properties of the hydrodynamic medium.  

As discussed in Section~\ref{sec:model_overview}, deviations of the plasma from local thermal equilibrium lead to modifications in the corresponding hadronic momentum distribution from Bose-Einstein or Fermi-Dirac. This ``viscous correction'' to the equilibrium distribution function is related to the magnitude of the bulk pressure. Its dependence on the hadron mass depends on the model used to calculate the viscous corrections. For the Grad model used in this work, this viscous correction increases with the hadron mass.\footnote{%
    We note that systematic studies of the mass dependence of bulk viscous corrections, and their effect on light nuclei production, could help differentiate between different models of viscous corrections.}
As a result, heavy particles such as protons, neutrons, and especially deuterons might be expected to have a higher sensitivity to bulk viscosity, compared to the majority of produced hadrons. Despite this, there has been no systematic study of the role of bulk viscosity in the production of deuterons until now.

In Fig.~\ref{fig:MAP}, we investigate the relative importance of bulk viscosity by comparing the yield of each particle (solid line) to the case where the bulk viscous correction at particlization has been set by hand to zero for all particles (dotted line).  One can see that the importance of the viscous correction indeed increases significantly with mass, and that the yield of deuterons is affected much more than that of lighter hadrons; generically the identified hadron multiplicity gets enhanced whereas the mean $p_T$ gets reduced by the bulk viscous correction.  Note that, while the distribution of deuterons at particlization does not have a strong effect on the final deuteron distribution (see discussion in Section~\ref{sec:model_comparisons}), any change in the distribution of protons and pions subsequently feeds down to the deuteron.

%
%

%

\begin{table}[tb]
\centering
\begin{tabular}{c|c|c}
\hline
parameter & min. & max  \\ 
\hline\hline
$(\zeta/s)_{\rm max}$ & 0.03 & 0.15 \\
\hline
$k$ & 0.3 & 2 \\
\hline
$w$ [fm] & 0.5 & 1.5 \\
\hline
\end{tabular}
\caption{Range of model parameters used to produce the model predictions in this section. Note that the highest bulk viscosity to entropy $\zeta/s$ attained in the fluid before switching (at temperature $\Tsw{} = 0.136$ GeV) ranges from $0.029\lesssim (\zeta/s)(\Tsw) \lesssim 0.143$; this is slightly smaller than the nominal maximum value $(\zeta/s)_{\rm max}$ defined at temperature $T_{\zeta, c} =  0.12$ GeV  (see Table~\ref{tab:table_params} for the value of the model parameters). 
}
\label{tab:table_prior}
\end{table}

\begin{figure}[tb] 
\centering

    \begin{tikzpicture}
    \node (img1) {\includegraphics[width=0.9\linewidth]{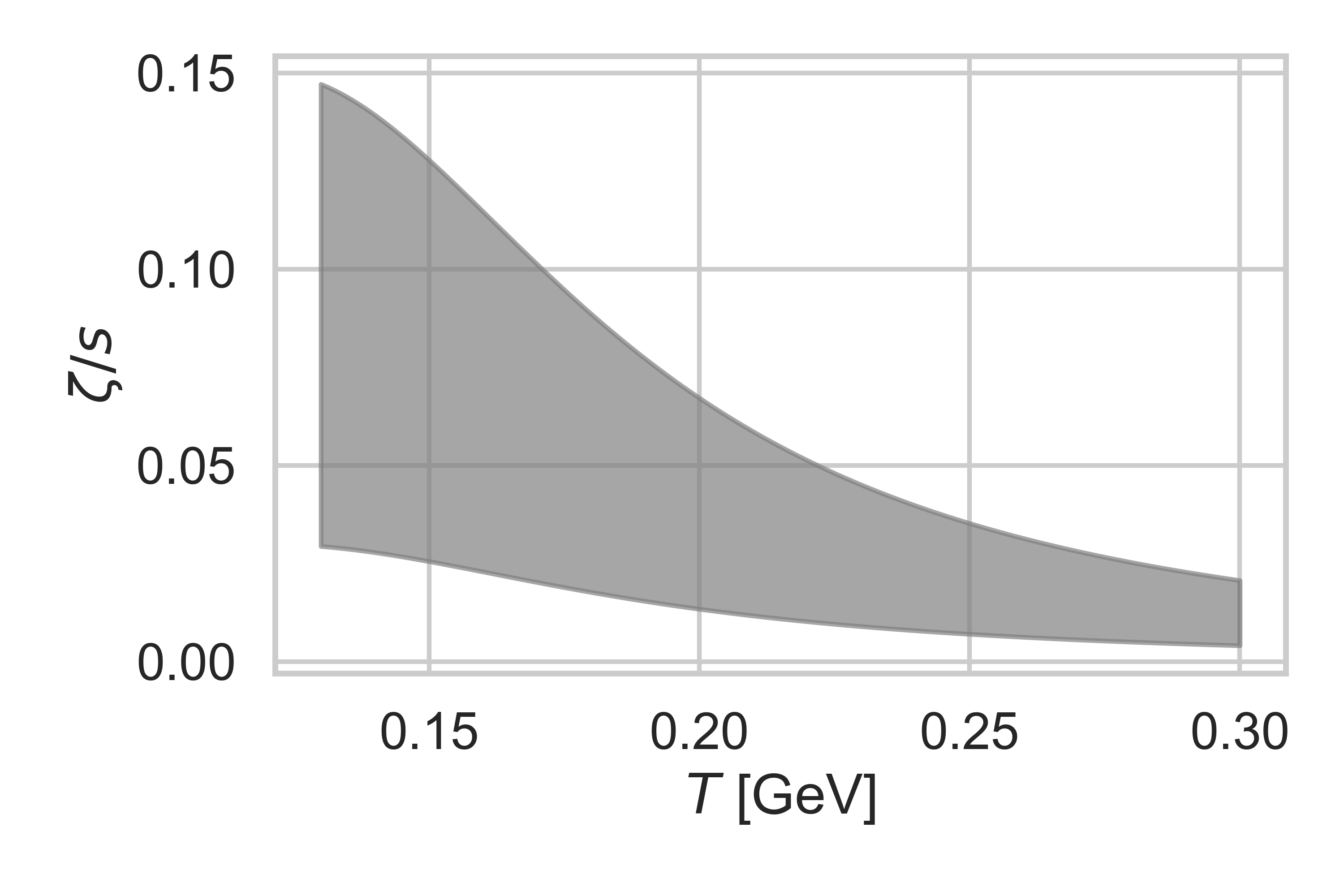}};
    \node[above right = -2.0cm and -3cm of img1] (img2) {\includegraphics[width=0.2\linewidth]{plots/jetscape_logo.jpg}};
    \end{tikzpicture}

\caption{The allowed range of specific bulk viscosity permitted by the chosen Bayesian prior. Only the magnitude was varied, while the shape parameters were held fixed by the values in Table~\ref{tab:table_params}. }
\label{fig:bulk_prior}
\end{figure}

To better understand the role of bulk viscosity on deuteron production, we proceed with a Bayesian analysis with three parameters of interest. Two parameters, $k$ and $w$ enter the initial conditions via the \trento{} model. The parameter $k \equiv 1/\sigma_k^2$ controls the magnitude of the fluctuations of the deposited energy in each nucleon-nucleon collision. The parameter $w$, referred to as the ``nucleon width'', controls the transverse radius of the nucleons in \trento{}, and defines the transverse size of deposited energy given a nucleon-nucleon collision. Both of these parameters largely control the homogeneity of the initially-deposited energy density. 
Finally, we vary the magnitude of the specific bulk viscosity at its peak value, $(\zeta/s)_{\rm max}$. The temperature dependence of bulk viscosity in this work is assumed to have a skewed-Cauchy form as in Ref.~\cite[Fig.~1]{JETSCAPE:2020mzn}. The priors for each parameter are assumed to be uniform within the ranges listed in Table~\ref{tab:table_prior}. The range of temperature-dependent specific bulk viscosities spanned by the prior for $(\zeta/s)_{\rm max}$ is shown in Fig.~\ref{fig:bulk_prior}. For deuteron production, we use the ``Coalescence only'' model described in Sections~\ref{sec:model_overview} and \ref{sec:model_comparisons}.

\begin{figure}[tb]
  \centering
  
    \begin{tikzpicture}
    \node (img1) {\includegraphics[width=0.9\linewidth]{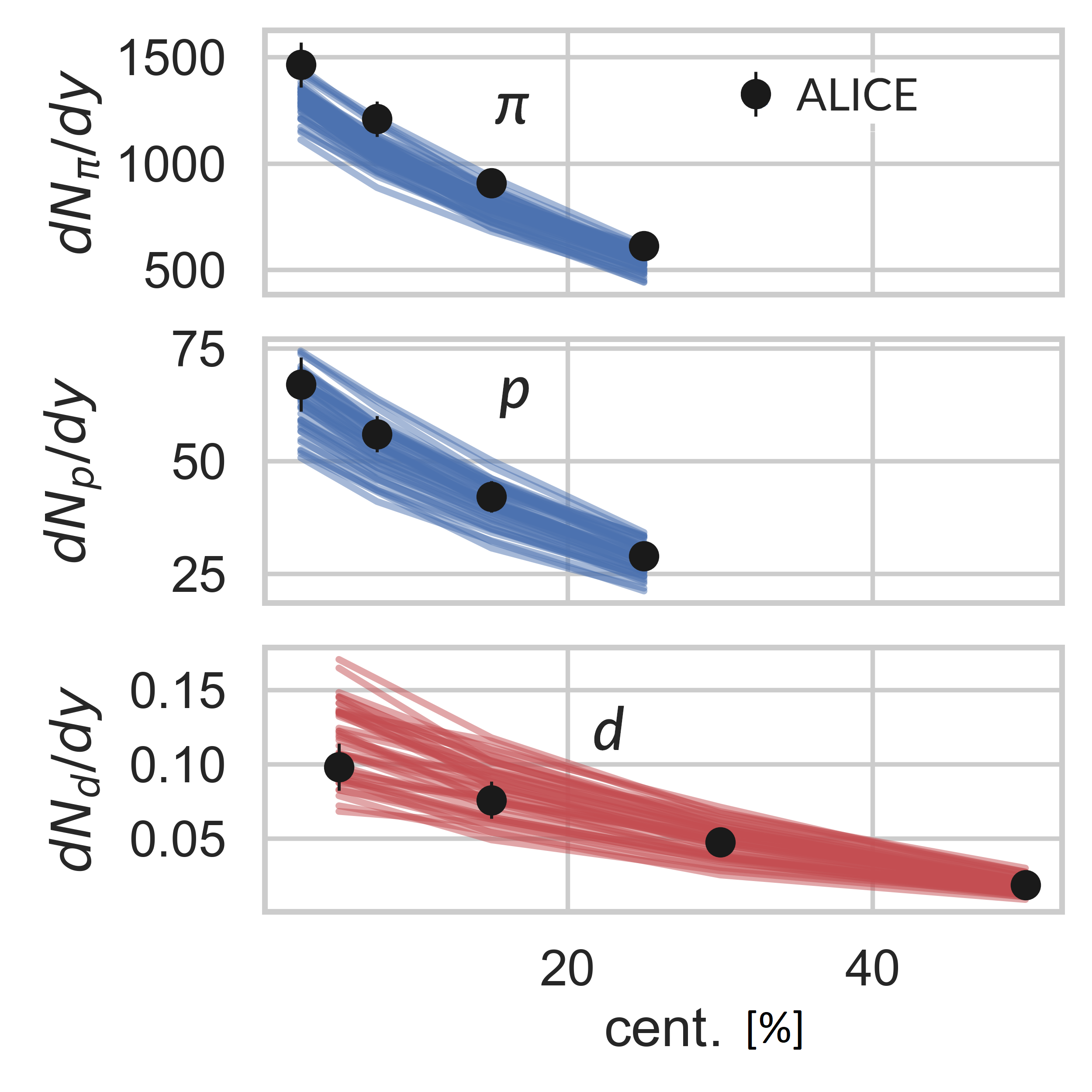}};
    \node[above right = -4cm and -2.5cm of img1] (img2) {\includegraphics[width=0.2\linewidth]{plots/jetscape_logo.jpg}};
    \end{tikzpicture}
    \begin{tikzpicture}
    \node (img1) {\includegraphics[width=0.9\linewidth]{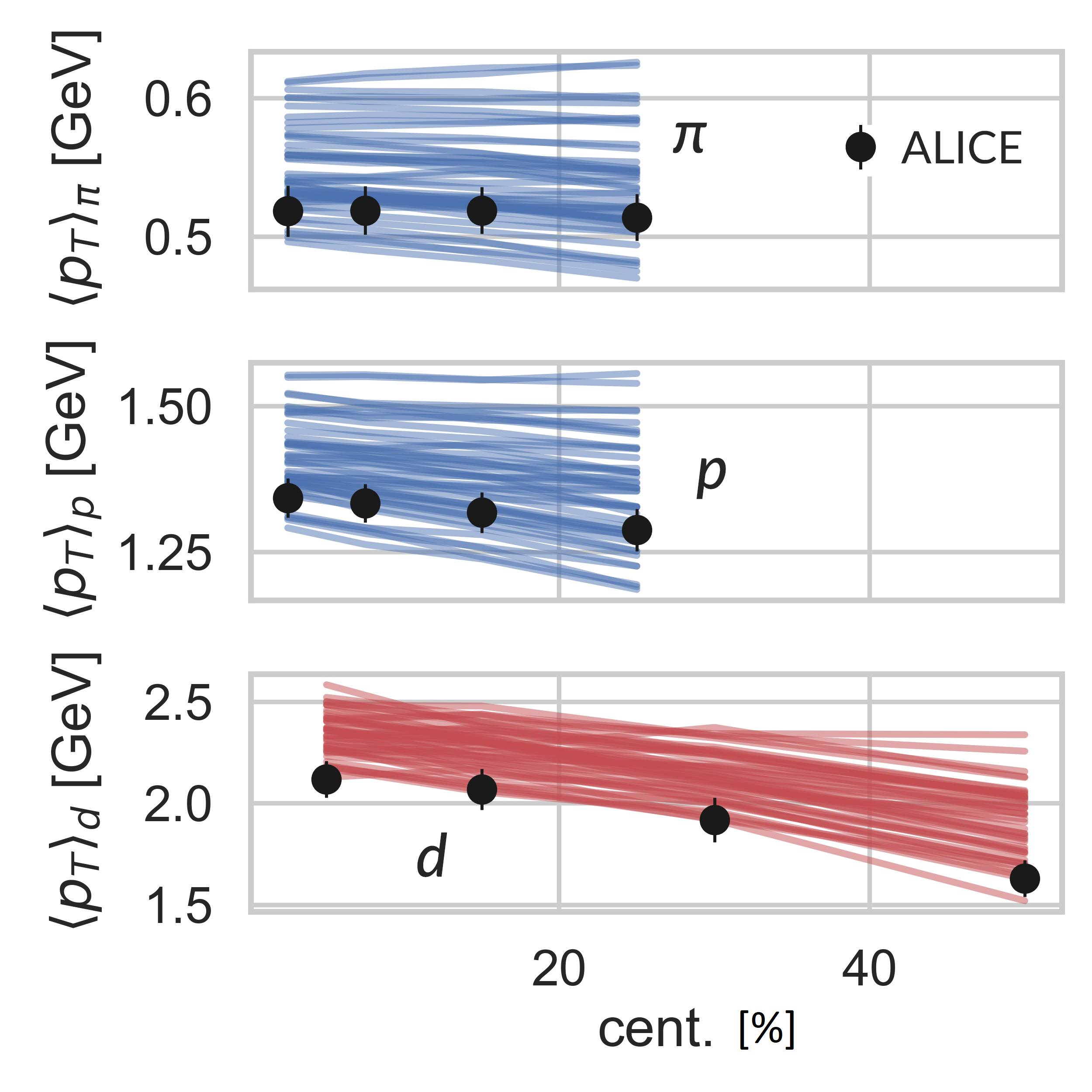}};
    \node[above right = -4cm and -2.5cm of img1] (img2) {\includegraphics[width=0.2\linewidth]{plots/jetscape_logo.jpg}};
    \end{tikzpicture}
    
  \caption{Prior predictive distributions for the particle yields (top) and mean $p_T$ (bottom), given by model predictions at the 45 points sampled uniformly in the prior volume. These points form the Gaussian process design. ALICE data from Refs~\cite{ALICE:2013mez,ALICE:2015wav}.
  \label{fig:prior_predict}}
\end{figure}

Forty-five design points were sampled from the prior using a Latin hypercube design \cite{santner2003design}. Combined with the fixed values of all remaining model parameters, the model's prediction  (given by the model's prior predictive distribution) for these 45 samples of $k$, $w$ and $(\zeta/s)_{\rm max}$ are shown in Fig.~\ref{fig:prior_predict} for the distributions of pion, proton and deuteron yields and mean $p_T$ in Pb-Pb $\sqrts{}=2.76$ TeV collisions.

We perform the Bayesian inference along the lines of previous works~\cite{Petersen:2010zt, Novak:2013bqa, Pratt:2015zsa, Sangaline:2015isa, Bernhard:2016tnd, Bernhard:2015hxa, Bernhard:2019bmu, JETSCAPE:2020mzn, JETSCAPE:2020shq}: For each observable, an emulator is used to interpolate the model's results between the parameter point samples.
The major difference with previous work is that we use a more sophisticated Gaussian process emulator.\footnote{Rather than performing a linear dimensionality reduction of the model outputs, such as Principal Component Analysis, we train a \textit{multitask} Gaussian process regression \cite{10.5555/2981562.2981582} as the model surrogate. If two observables are labeled by $i$ and $j$, and two vectors of model parameters labeled by ${ \boldsymbol x}$ and ${\boldsymbol x}'$, then the multitask kernel function is given by 
\begin{equation}
    k_{ij}({ \boldsymbol x}, { \boldsymbol x}') = k({ \boldsymbol x}, { \boldsymbol x}')k_{\rm task}(i,j)
\end{equation}
where $k( \boldsymbol{x}, \boldsymbol{x}')$ plays the role of the usual kernel function, and $k_{\rm task}(i,j)$ is a kernel function which models the correlations among different outputs. }

\begin{figure}[b]
  \centering
  
    \begin{tikzpicture}
    \node (img1) {\includegraphics[width=0.95\linewidth]{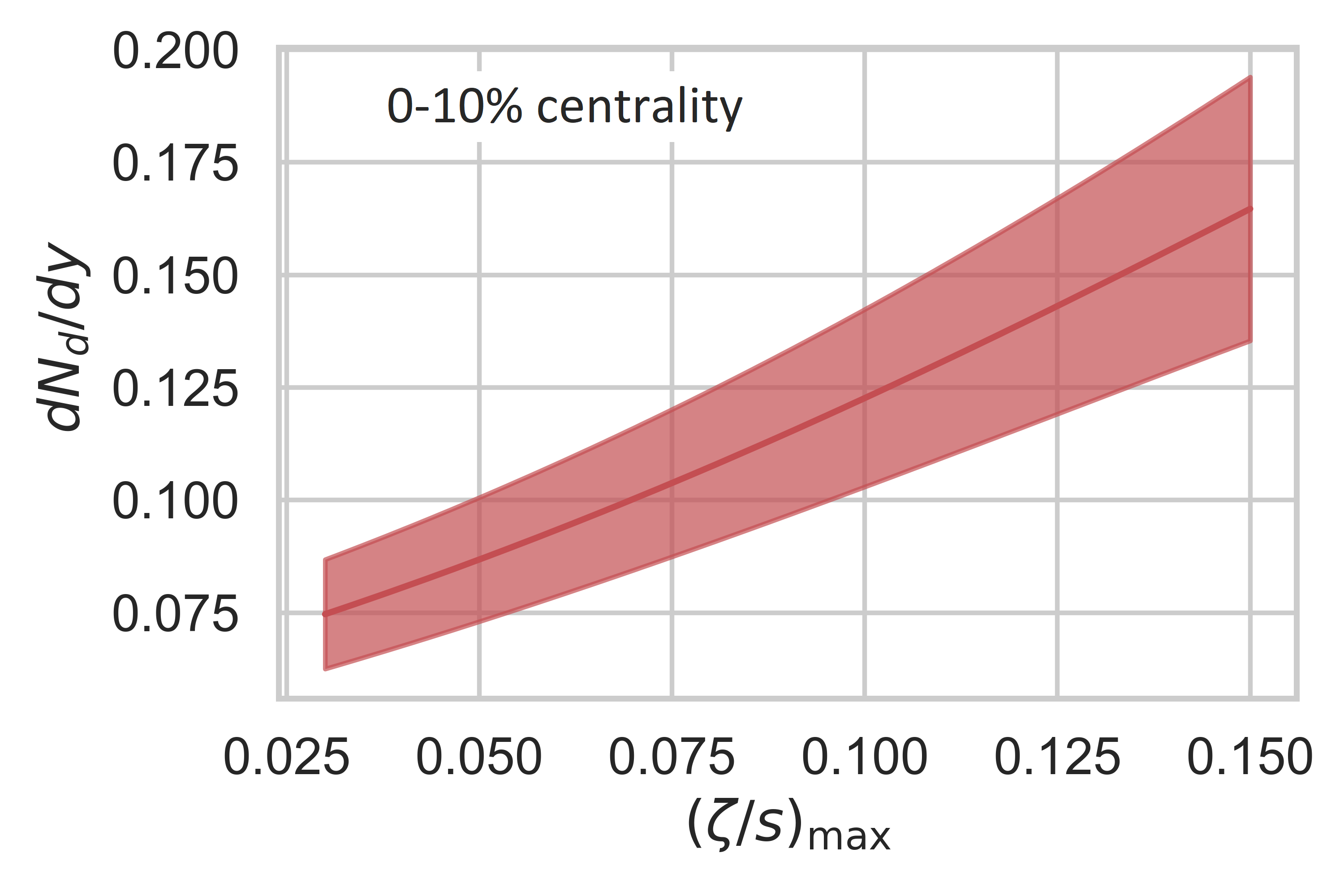}};
    \node[above right = -2.7cm and -6.5cm of img1] (img2) {\includegraphics[width=0.2\linewidth]{plots/jetscape_logo.jpg}};
    \end{tikzpicture}

  \caption{The response of the deuteron yield $dN_d/dy$  to the change of the magnitude of the specific bulk viscosity $(\zeta/s)_{\rm max}$, for Pb-Pb $\sqrts{}=2.76$~TeV 0--10\% centrality.}
  \label{fig:deuteron_zeta}
\end{figure}

To illustrate the sensitivity of deuteron observables to the magnitude of the bulk viscosity, we fix the \trento{} fluctuation $k$ and nucleon width $w$ to the midpoints of their prior, and plot the emulated response of the deuteron yield to changes in the specific bulk viscosity $(\zeta/s)_{\rm max}$. This is shown in Fig.~\ref{fig:deuteron_zeta}, and we see that the deuteron yield indeed shows strong sensitivity to the magnitude of bulk viscosity. 

\begin{figure}[tb]
  \centering
  
    \begin{tikzpicture}
    \node (img1) {\includegraphics[width=\linewidth]{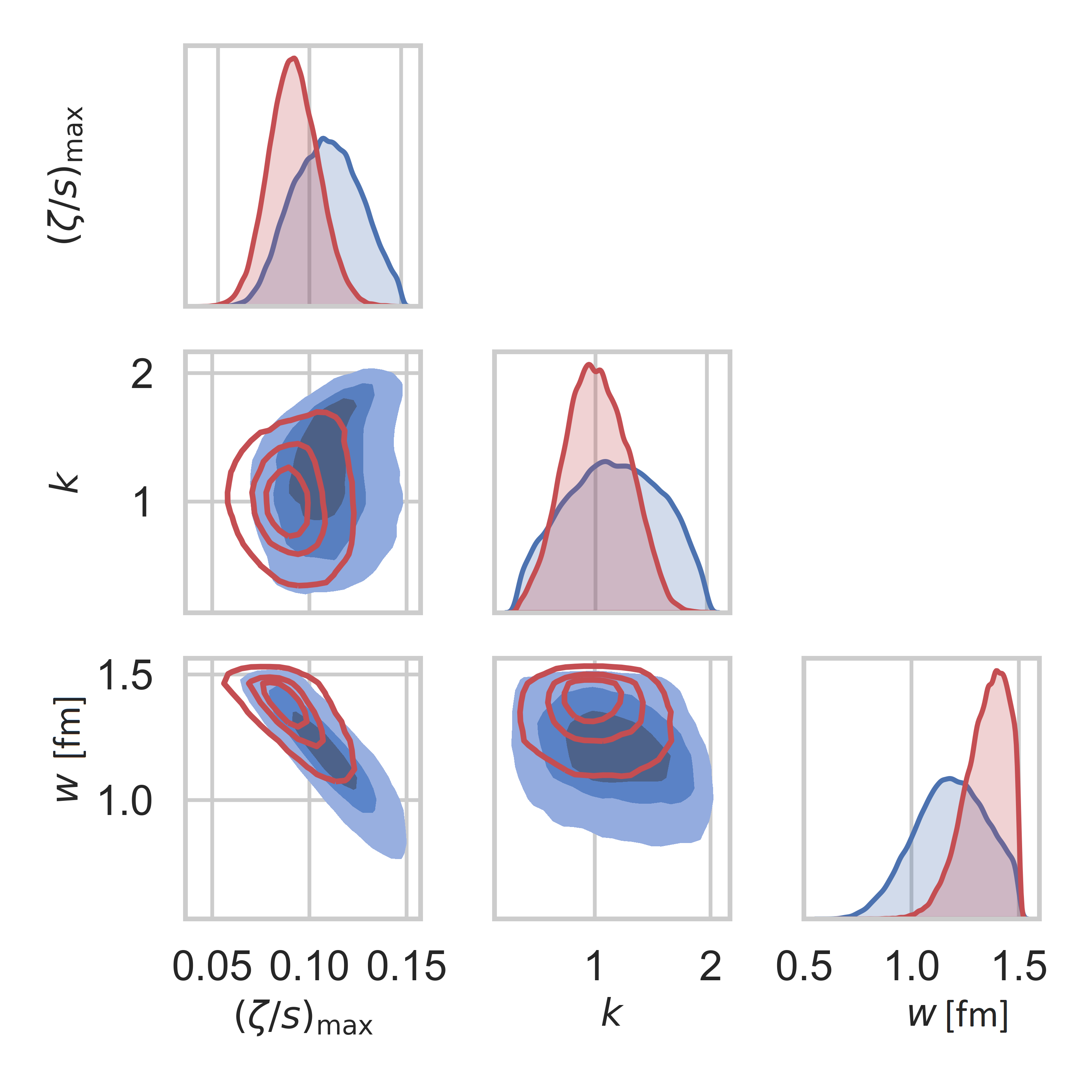}};
    \node[above right = -1.7cm and -5.2cm of img1] (img2) {\includegraphics[width=0.2\linewidth]{plots/jetscape_logo.jpg}};
    \end{tikzpicture}

  \caption{The posterior density of single (diagonal) and joint (off diagonal) marginal posterior distributions of the three model parameters, calibrated to pion and proton observables (shaded blue) or pion, proton and deuteron observables (unshaded red). }
  \label{fig:corner}
\end{figure}

In Fig.~\ref{fig:corner} we show the effect of adding deuteron observables on constraining the nucleon width $w$, the fluctuation parameter $k$ and the maximum of bulk viscosity $(\zeta/s)_{\rm max}$. We see that the deuteron's dependence on bulk viscosity modifies the value of $(\zeta/s)(T)$ that is in best agreement with measurements. Moreover, Fig.~\ref{fig:corner} shows how this change in bulk viscosity correlates with changes in the preferred value of the initial condition parameters.
For example, we see that including deuteron observables would favor a slightly smaller $(\zeta/s)_{\rm max}$ and a larger $w$; the former can be understood from the fact that simulations with parameters calibrated without deuteron observables overestimate the deuteron multiplicity (see Fig.~\ref{fig:3-way}a) whereas decreasing $(\zeta/s)_{\rm max}$ helps to reduce the tension (see Fig.~\ref{fig:deuteron_zeta}), and the latter can be understood by its anticorrelation with $w$. We do not intend these results to represent accurate calibrations of the three model parameters considered here, in part because we only explored a small subspace of the much larger parameter space considered in Ref.~\cite{JETSCAPE:2020mzn}, and in part because we only calibrated the three model parameters against a small set of observables. Nevertheless, Fig.~\ref{fig:corner} illustrates the sensitivity of deuterons to bulk viscosity and their potential for improving constraints on the properties of quark-gluon plasma.

\section{Summary}
\label{sec:conclusions}

We explored deuteron production in ultra-relativistic Pb+Pb collisions at $\sqrts$ of 2.76 TeV and 5.02 TeV, two collision systems where recent deuteron measurements are available. For this purpose we employed a multistage approach (hydrodynamics + hadronic afterburner) tuned to reproduce the yields, mean transverse momenta, and flow of pions, kaons, and protons in Au-Au collisions at $\sqrts{}=0.2$~TeV and Pb-Pb collisions at $\sqrts{}=2.76$~TeV; no deuteron observables were used for tuning. Three different models of deuteron production were tested -- ``Transport only'', ``Cooper-Frye + Transport'', ``Coalescence only'' (Section~\ref{sec:model_comparisons}). Overall, all three models produce rather similar results. At 2.76 TeV they reproduce the centrality dependence of deuteron $\langle p_T \rangle$ and $v_2(p_T)$ within error bars, while the deuteron yields are overestimated in central collisions but reproduced well in more peripheral ones. It is  possible that a more realistic deuteron wave function might affect the centrality dependence and lead to improvement in central collisions but we have not checked this.

Our predictions for deuteron flow at 5.02 TeV had been confronted with experimental data in Ref.~\cite{ALICE:2020chv}, with good overall agreement. This was expected, since the deuteron flow is not much different at 2.76 and 5.02 TeV and the model reproduced the deuteron $v_2(p_T)$ at 2.76 TeV very well. The data for deuteron yields at 5.02 TeV are not yet published by ALICE -- although analogously to 2.76 TeV, it would not be surprising that our prediction overestimates the yield in central collisions, reproduces it in peripheral collisions, and reproduces the $\langle p_T \rangle$ precisely. Despite the above tension with 2.76 TeV measurements in central collisions, we expect our prediction for the ratio of measured yields at 5.02 TeV and 2.76 TeV to be more robust.

The main conclusion of this study is that deuteron observables are particularly sensitive to bulk viscosity. 
We have seen in Fig.~\ref{fig:MAP} that when bulk viscous corrections change the proton and neutron yield by 20--25\%, the deuteron yield can be changed by as much as 50\%. 
While the quantitative values for the bulk viscous corrections quoted above are quite large --- and might be even pushing the multistage model to its limits~\cite{Byres:2019xld,Bemfica:2020xym,Chiu:2021muk,plumberg2021causality} --- the stronger dependence on bulk viscosity of deuterons compared to protons should be robust.

The fact that deuterons are sensitive to the bulk viscous corrections has an interesting implication: proton femtoscopic radii should also be sensitive to the bulk viscosity. Indeed, a relation between proton femtoscopic radii and coalescence has been explicitly demonstrated recently \cite{Blum:2019suo}.

The overall dependence of light nuclear observables on bulk viscosity could be used to improve constraints on this transport coefficient, as discussed in Section~\ref{sec:sensitivity_bulk}. We have provided a preliminary constraint in Fig.~\ref{fig:corner}; a more robust constraint will require a better understanding of the bulk viscosity in heavy ion collisions, in particular viscous corrections at the transition between hydrodynamics and transport.

\acknowledgements

This work was supported in part by the National Science Foundation (NSF) within the framework of the JETSCAPE collaboration, under grant number OAC-2004571 (CSSI:X-SCAPE). It was also supported under ACI-1550172 (Y.C. and G.R.), ACI-1550221 (R.J.F., F.G., and M.K.), ACI-1550223 (D.E., U.H., L.D., and D.L.), ACI-1550225 (S.A.B., T.D., W.F., R.W.), ACI-1550228 (R.E., B.J., P.J., J.M., X.-N.W.), and ACI-1550300 (S.C., A.K., A.M., C.N., A.S., J.P., L.S., C.Si., R.A.S. and G.V.); by PHY-1516590 and PHY-1812431 (R.J.F., F.G., M.K.), by PHY-2012922 (C.S.); it was supported in part by NSF CSSI grant number \rm{OAC-2004601} (BAND; D.L. and U.H.); it was supported in part by the US Department of Energy, Office of Science, Office of Nuclear Physics under grant numbers \rm{DE-AC02-05CH11231} (X.-N.W.), \rm{DE-FG02-00ER41132} (D.O.), \rm{DE-AC52-07NA27344} (A.A., R.A.S.), \rm{DE-SC0013460} (S.C., A.K., A.M., C.S. and C.Si.), \rm{DE-SC0004286} (L.D., D.E., U.H. and D. L.), \rm{DE-FG02-92ER40713} (J.P.), \rm{DE-FG02-05ER41367} (T.D., J.-F.P., D.S. and S.A.B.), \rm{DE-SC0021969} (C.S.), and \rm{DE-FG05-92ER40712} (L.K., A.M., J.V.), and contract number \rm{DE-SC0012704} (B.S. and C.S.). The work was also supported in part by the National Science Foundation of China (NSFC) under grant numbers 11935007, 11861131009 and 11890714 (Y.H.), by the Natural Sciences and Engineering Research Council of Canada (C.G., M.H., S.J., and G.V.), by the Office of the Vice President for Research (OVPR) at Wayne State University (Y.T.), by the S\~{a}o Paulo Research Foundation (FAPESP) under projects 2016/24029-6, 2017/05685-2, 2018/24720-6 and 2021/08465-9 (A. L. and  M.L.), and by the University of California, Berkeley - Central China Normal University Collaboration Grant (W.K.). U.H. would like to acknowledge support by the Alexander von Humboldt Foundation through a Humboldt Research Award. C.S. acknowledges a DOE Office of Science Early Career Award. Allocation of super-computing resources (Project: PHY180035) were obtained in part through the Extreme Science and Engineering Discovery Environment (XSEDE), which is supported by National Science Foundation grant number ACI-1548562. Calculations were performed in part on Stampede2 compute nodes, generously funded by the National Science Foundation (NSF) through award ACI-1134872, within the Texas Advanced Computing Center (TACC) at the University of Texas at Austin \cite{TACC}, and in part on the Ohio Supercomputer \cite{OhioSupercomputerCenter1987} (Project PAS0254). Computations were also carried out on the Wayne State Grid funded by the Wayne State OVPR.  Data storage was provided in part by the OSIRIS project supported by the National Science Foundation under grant number OAC-1541335.

\bibliography{inspire,noninspire}

\end{document}